\documentclass{appolb}
\usepackage{epsfig}

\begin{document}
\title{Cosmology of Multi-Singlet-Scalar Extensions of the Standard Model
\thanks{Presented at the XXXV International Conference of Theoretical Physics "Matter to the Deepest 2011"}%
}
\author{A. Drozd and B. Grzadkowski
\address{Institute of Theoretical Physics, University of Warsaw, 
Ho\.za 69, PL-00-681 Warsaw, Poland}
\and
Jos\'e Wudka
\address{Department of Physics, University of California,
Riverside CA 92521-0413, USA}
}
\maketitle
\begin{abstract}
An extension of the Standard Model by extra scalar singlets was considered. Theoretical (unitarity, vacuum stability, triviality) and cosmological (dark matter relic abundance, direct detection experiments, constraints on dark matter self-interaction) constraints were discussed.
\end{abstract}
  
\section{Introduction}
The Standard Model (SM) of electro-weak interactions does not provide a viable candidate for Dark Matter (DM)~\cite{Jarosik:2010iu}. To unravel the DM mystery we need to consider its extensions. Among many existing proposals  there is the simplest DM model, constructed by adding a real singlet scalar field to the SM particles. We would like to propose a variation of the singlet scalar extension consisting of an adition of the $N$ extra scalar singlets transforming according to the fundamental representation of $O(N)$ symmetry group.

In this paper we confront our model with theoretical bounds (unitarity, stability, triviality) and cosmological data (the amount of DM in the Universe \cite{Jarosik:2010iu} and direct detection XENON 100 experiment \cite{Aprile:2011hi}). We also note, that conventional collisionless cold dark matter has problems explaining the observed structure of galaxies and significant self-interaction rate provides possible cure (see section \ref{sec:SIDM} and \cite{Spergel:1999mh}).

\section{Model and Theoretical Constraints}
We shall consider the SM extended by addition of $N$ scalars $\varphi_i$ which are singlets under the SM gauge group $SU(3)\times SU(2)\times U(1)$. We assume that the $N$-component vector $\vec{\varphi}$ transforms according to the fundamental representation of $O(N)$ - an exact symmetry of the model. All SM fields are singlets under $O(N)$. For the purpose of providing a DM candidate we impose an additional $Z_2$ symmetry, under which $\vec{\varphi}$ is odd: $\vec{\varphi}\to-\vec{\varphi}$.
The most general, symmetric and renormalizable potential is:
\begin{eqnarray}
V(H, \vec{\varphi}) = 
- \mu_{H}^2 H^{\dagger} H 
+ \lambda_{H} (H^{\dagger} H)^2 
+ 1/2 \mu_{\varphi}^2 \vec{\varphi}^2
+ 1/4! \lambda_{\varphi} \left(\vec{\varphi}^2 \right)^2
 +  \lambda_{x} H^{\dagger} H \vec{\varphi}^2
 \label{pot}
\end{eqnarray}
The Higgs field is a SM doublet with a vacuum expectation value (VEV), $\langle H \rangle = v/\sqrt{2}$ for $v = 246$ GeV. After the symmetry breaking  the Higgs-boson mass is $m_h^2 = - \mu_{H}^2 + 3 \lambda_{H} v^2 =2 \mu_{H}^2$, as in the SM. The singlet's masses also get a VEV contribution, $m_{\varphi}^2 = \mu_{\varphi}^2 + \lambda_{x} v^2$.

Singlets do not develop a VEV - we would like them to be stable, so the $Z_2$ must be unbroken with $\mu_{\varphi}^2 \geq 0$. That implies 
\begin{equation} 
m_{\varphi}^2 \geq \lambda_{x} v^2
\label{mphi_bound}
\end{equation} which is a significant constraint - light scalars ($m_{\varphi} \ll v$) must couple very weakly to the SM. Quartic couplings are constrained by the unitarity arguments for longitudinal W boson- and scalar-scattering \cite{Lee:1977eg}, \cite{Cynolter:2004cq}: 
\begin{equation}
m_{H}^2< (8 \pi)/3 v^2, \;  \lambda_{\varphi} < 8 \pi, \; |\lambda_x| < 4 \pi \label{unitarity}
\end{equation}

Tree level vacuum stability of the scalar potential (\ref{pot}) implies that either all the quartic couplings are positive: 
$\lambda_{H}, \lambda_{\varphi}, \lambda_x > 0 $
or $\lambda_x$ is negative and  
$\lambda_x^2 <  \lambda_{\varphi}\lambda_{H}/6  =  \lambda_{\varphi} m_h^2/(12 v^2)$ \cite{Cynolter:2004cq}.

\subsection{Triviality}
We require that the during the renormalization group equation (RGE) running, the quartic couplings $\lambda_H, \lambda_x, \lambda_{\varphi}$ remain finite up to the cut-off scale $\Lambda$. That implies limits on the model parameters, in particular an upper limit on $m_{h}$ (so called 'triviality bound').

The RGE for running couplings in our model can be found in \cite{OD-MSc-thesis}. We solve those equations with initial conditions:
$ \lambda_H(\mu = m_W)~=~\lambda_{H\, 0}$, 
$\lambda_{x}(\mu~=~m_W)~=~\lambda _{x \, 0}$ and
$\lambda_{\varphi}(\mu = m_W) = \lambda _{\varphi \, 0} $.

\begin{figure}[tp]
  \centering
\includegraphics[width = 6 cm, height = 5.1 cm]{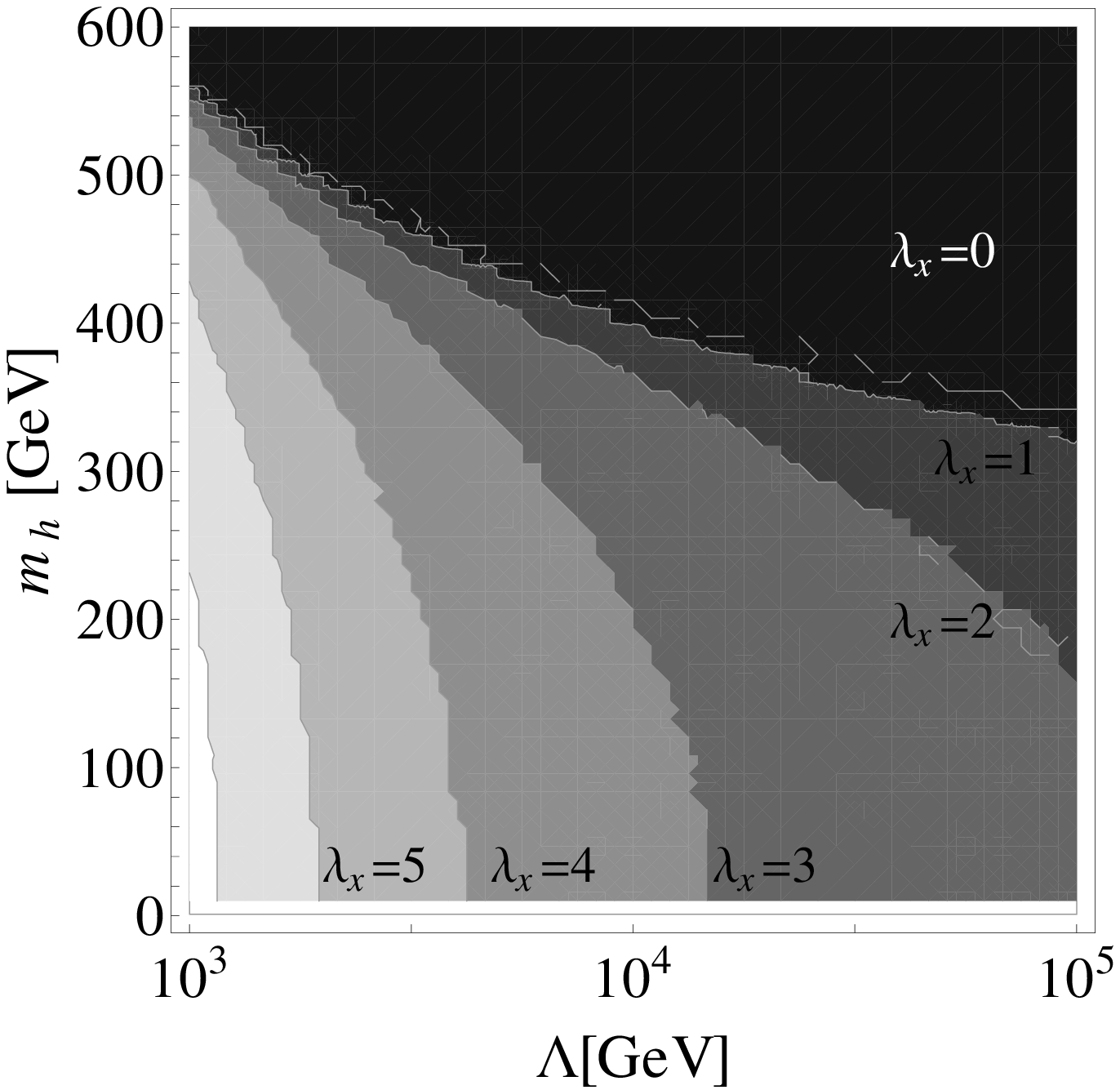}
\includegraphics[width = 6 cm, height = 5 cm]{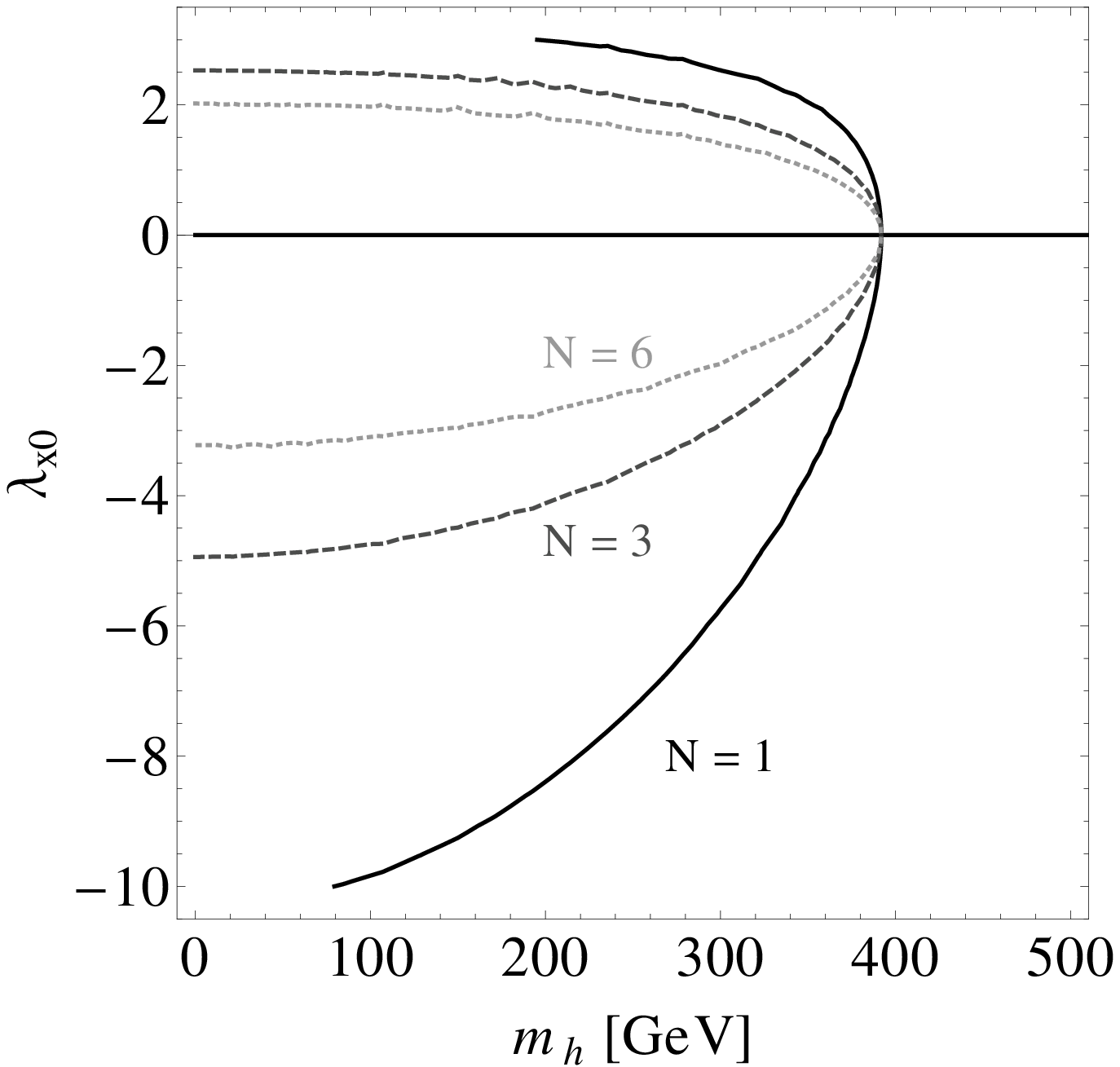}
  \caption{LEFT: Triviality upper bound for the Higgs boson as a function of the cut-off $\Lambda$ for different $\lambda_{x 0}$, $N=1$ and $\lambda_{\varphi 0} = 0.1$. Regions above each curve are forbidden.
RIGHT: Regions (between curves of the same type) allowed for $\lambda_{x 0}$ by triviality  for $\Lambda = 10^4 \, \mathrm{ GeV}$, $\lambda_{\varphi 0} = 0.5$.}
  \label{triv_N}
\end{figure}

We assume that for a given $\Lambda$ there is no pole in the evolution of the scalar quartic couplings at energies below $\Lambda$. That gives us constraints in the $(m_h,\Lambda)$ plane depending on initial parameters $ \lambda _{x \, 0}, \lambda _{\varphi \, 0}$ and $N$ - see the left panel of fig.\ref{triv_N}. Note that the allowed region shrinks as $\lambda_{x 0}$ grows and the upper bound on $m_h$ is getting lower. There is also an asymmetry between the negative and positive branch of $\lambda_{x0}$ initial conditions, see the right panel of fig.\ref{triv_N}

\section{Experimental Constraints}
\subsection{DM relic abundance - Cold Dark Matter (CDM)}

Stability of the singlets $\vec{\varphi}$ makes them good DM candidates. To calculate DM relic abundance in a case of compound DM we need in general a set of Boltzmann equations for all component. Because of the $O(N)$ symmetry it actually simplifies to one equation \cite{kolb}:
\begin{eqnarray}
 \frac{d f}{d T}  = \frac{ \left<\sigma v\right>}{K} (f^2 - f_{EQ}^2), \,\,\,\,\, K(T) =\sqrt{ \frac{4 \pi^3 g_\star(T)}{45 m_{Pl}^2} }
 \label{b_eq}
\end{eqnarray}
where $f \equiv n/T^3$, $n$ is the number density of DM, $f_{EQ}$ is the equilibrium distribution, $ g_\star(T)$ - number of relativistic degrees of freedom, $m_{Pl}$ - Planck mass,  $\left<\sigma v\right>$ - the thermally averaged cross section for $DM + DM \to SM~+~SM$ processes \cite{Guo:2010hq}, \cite{Gondolo:1990dk}. Total abundance of DM, $\Omega_{\mathrm{DM}}$,  reads
\begin{eqnarray}
\Omega_{\mathrm{DM}}^N = \sum_{i} \Omega_{\mathrm{DM}}^{i} = N  \Omega_{\mathrm{DM}}^{1}
\label{omega_N}
\end{eqnarray}
where $\Omega_{\mathrm{DM}}^{i}$ is the dark matter relic density from $i$-th scalar field $\varphi_i$.

\begin{figure}[tp]
  \centering
\includegraphics[width= 6.2 cm]{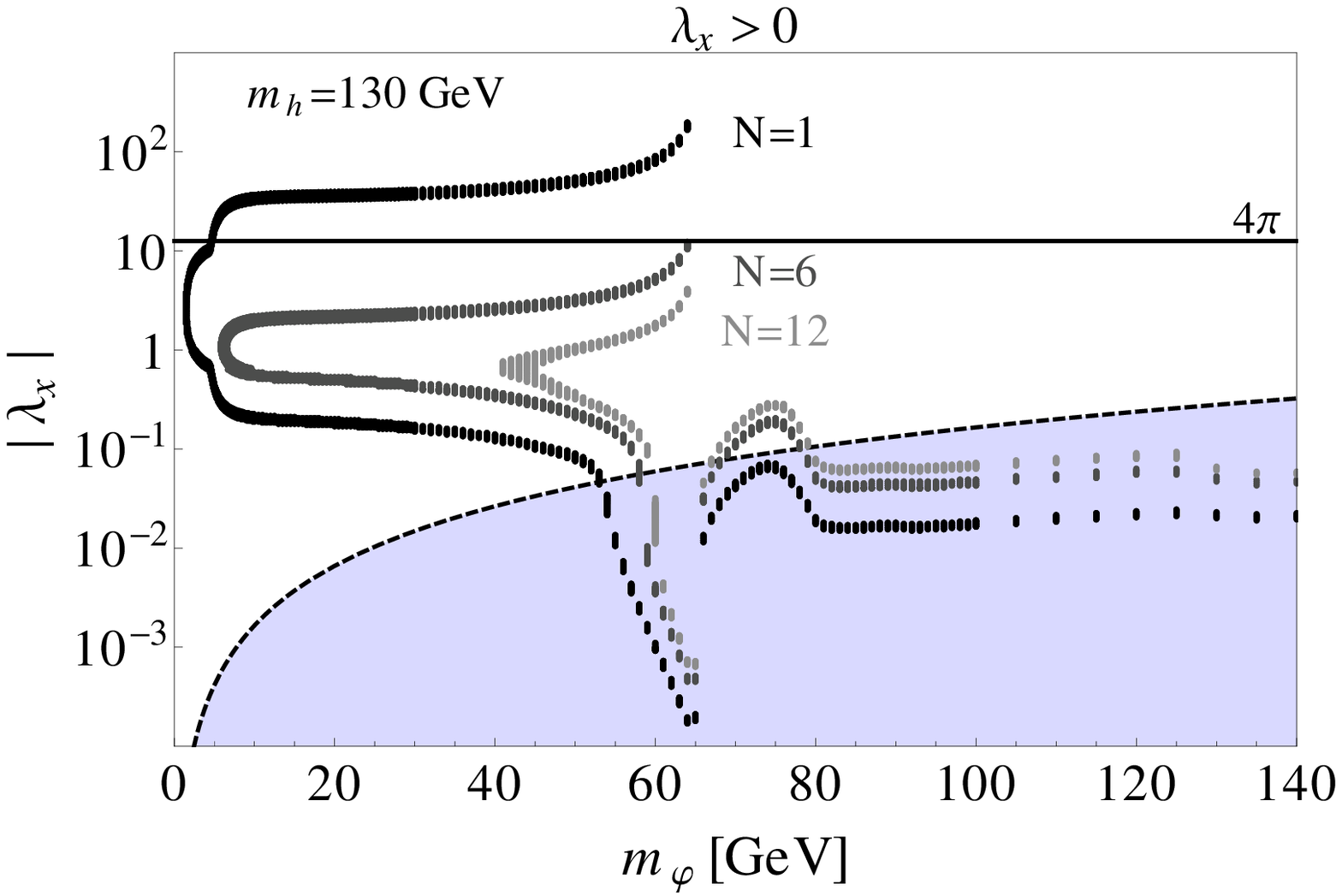}
\includegraphics[width= 6.2 cm]{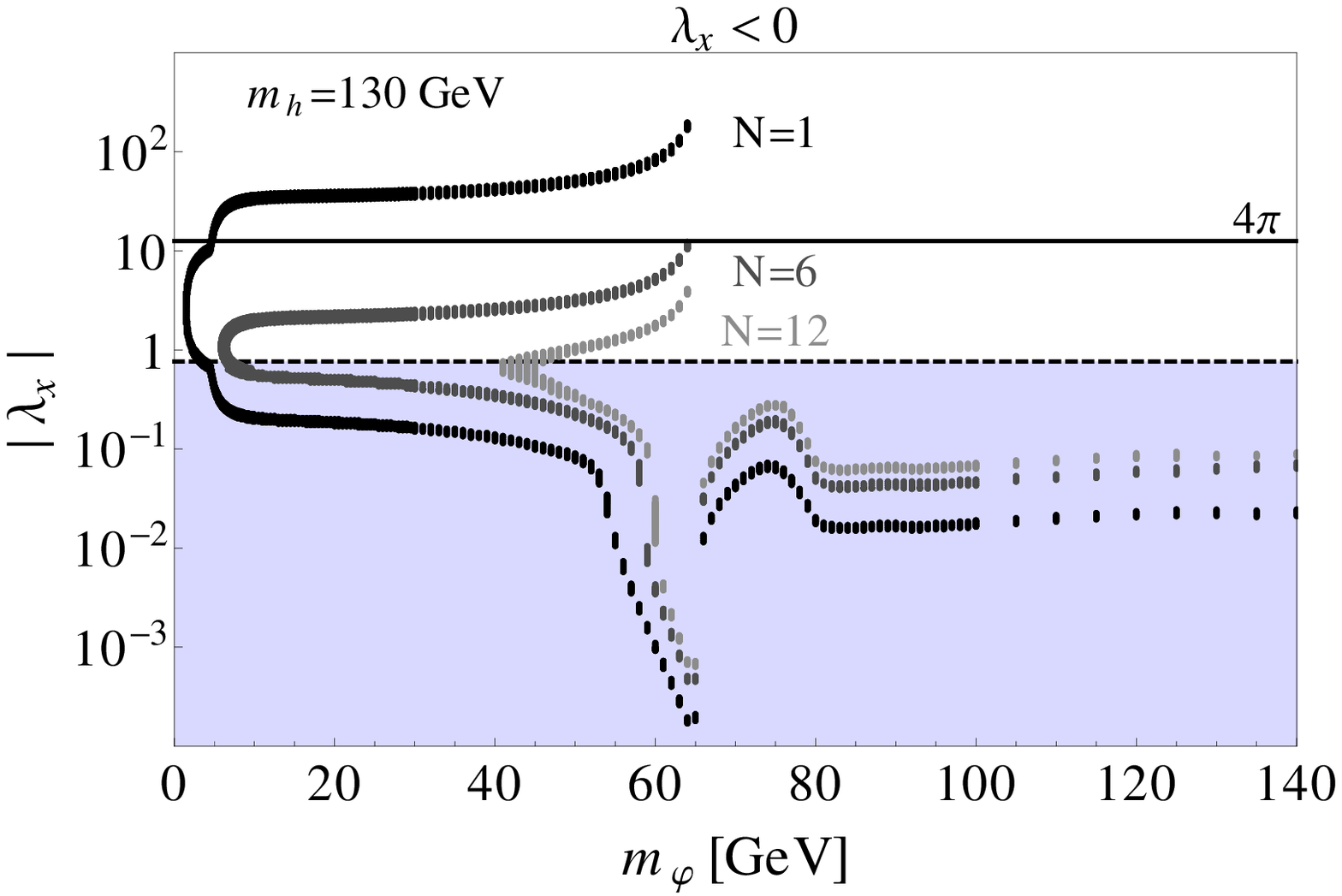}
  \caption{ $|\lambda_x|$ as a function of $m_{\varphi}$ for CDM, $m_{h}=~130 \mathrm{ GeV}$, $\lambda_x>0$ (left panel) and $\lambda_x<0$ (right panel). The dark gray, gray and light gray points correspond to  $N=~1,6$ and $12$, respectively. The blue areas in the left and right panels show regions allowed by the consistency condition (\ref{mphi_bound}) and the vacuum 
  stability for  $\lambda_{\varphi}=8\pi$, respectively. The thick black lines show the unitarity limit (\ref{unitarity}) $|\lambda_x|=4\pi$.}
  \label{phases}
\end{figure}

We solve (\ref{b_eq}) in the standard CDM case \cite{kolb} using MicrOMEGAs~\cite{Belanger:2008sj}. From the WMAP data \cite{Jarosik:2010iu} we know that $\Omega_{\mathrm{DM}}^{(exp)} h^2= 0.110 \pm 0.018$ (allowing for $3 \sigma$ uncertainty) and for a given choice of $N$, $m_h$ and $m_{\varphi}$ we seek $\lambda_x$ 
such that this constraint is satisfied, see fig. \ref{phases}. Note that in the vicinity of resonance, the annihilation cross section is enhanced ($m_h \sim 2 m_{\varphi}$), therefore $\lambda_x$ is suppressed to reach the desired DM abundance. 

\subsection{DM relic abundance - Feebly Interacting Dark Matter (FIDM)}

 In the CDM model we achieve equilibrium between certain particles and the SM species just to lose it ('freeze-out') while the Universe cools. What happens if the DM particles interact with SM so feebly ($\lambda_x~<~10^{-9}$), that equilibrium with the SM species is never achieved \cite{Yaguna:2011qn}?

\begin{figure}[tp]
  \centering
\includegraphics[width=6.1 cm, height = 4 cm]{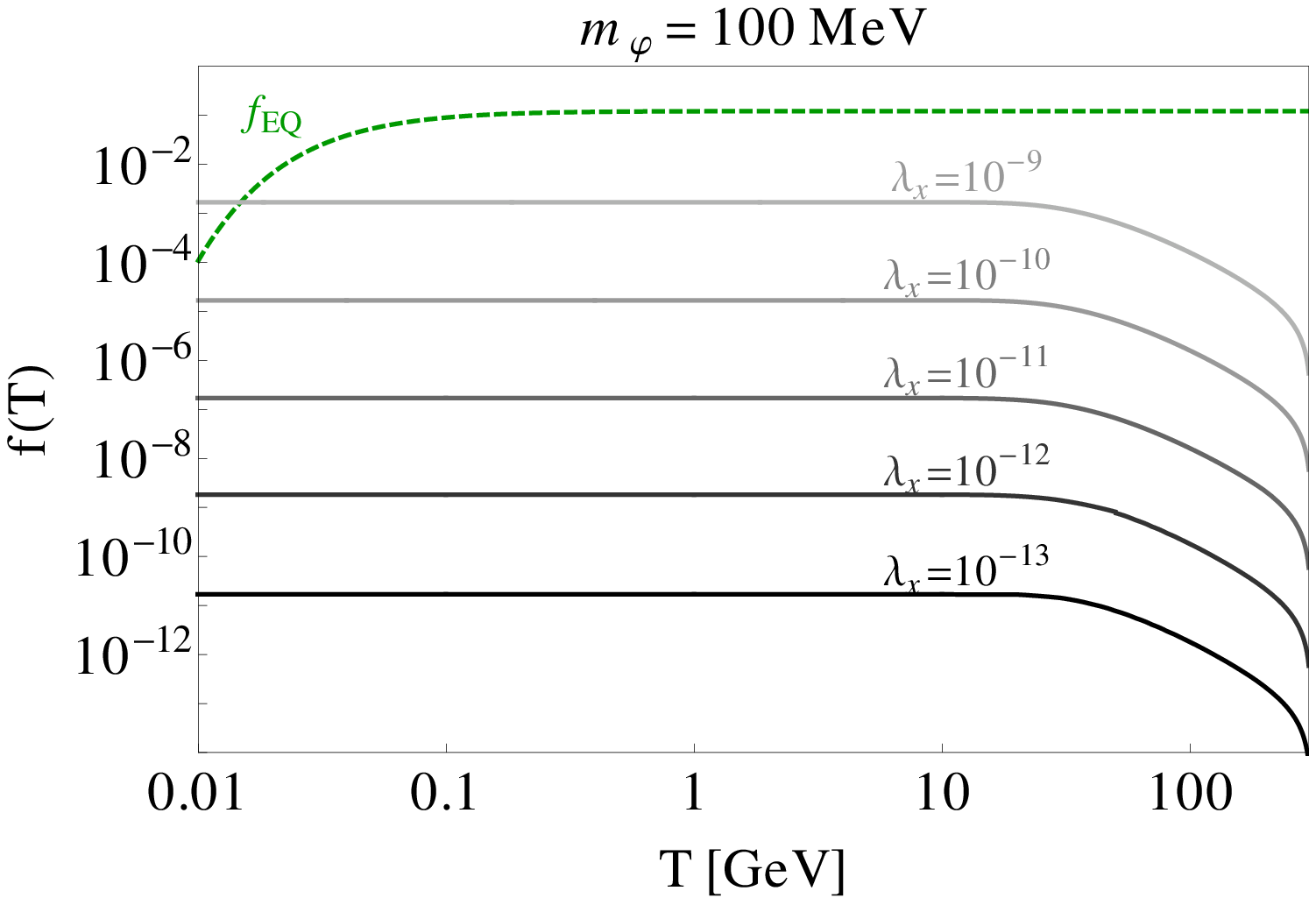}
\includegraphics[width=6.25 cm, height = 3.65 cm]{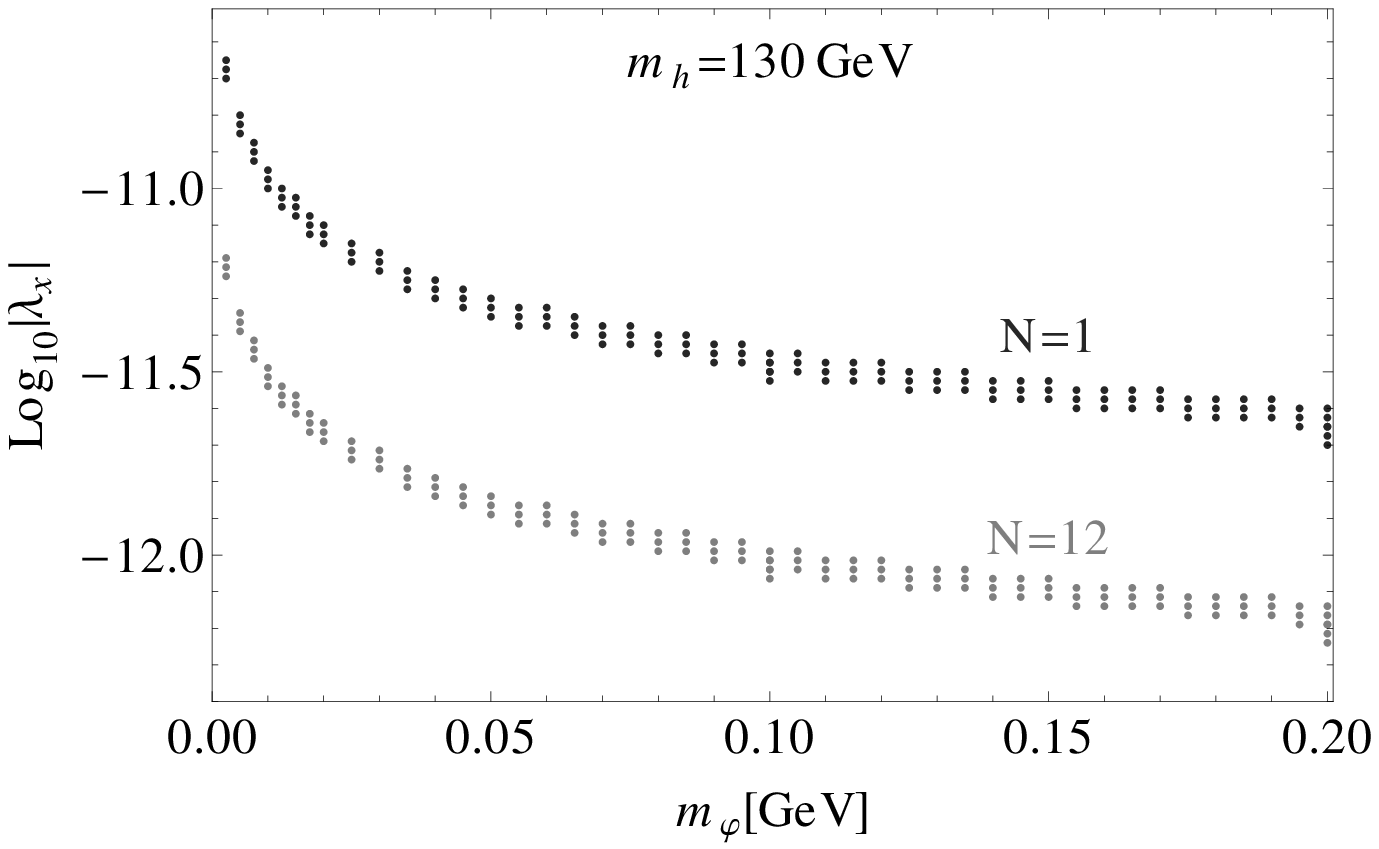}
  \caption{LEFT: Solutions to the Boltzman equation (\ref{b_eq}) in case of FIDM for a single scalar of mass $m_{\varphi} = 100$ MeV, $\lambda_x = 10^{-13}$ (bottom curve), $\dots$, $10^{-9}$ (top curve) for $m_h = 130$ GeV. Green dashed curve is the equilibrium distribution $f_{EQ}$. \newline
RIGHT:  FIDM solutions satisfying the relic abundance condition for $m_h = 130$ GeV, $N = 1$ (darker points), 12 (lighter).}
  \label{Boltzman}
\end{figure}

In the following we will assume the number density $f$ was negligible at the Big Bang: $\lim_{T \rightarrow \infty}{f(T)} = 0$. Having $f$ determined by the Boltzmann equation (\ref{b_eq}) (see fig. \ref{Boltzman}, left panel) one can get the DM relic abundance:
\begin{eqnarray}
\Omega_{\mathrm{DM}} h^2 = m_{\varphi} n_{\varphi}/\rho_{crit} = m_{\varphi} T_{\gamma}^3 f_{\varphi} / \rho_{crit}
\label{om-fidm}
\end{eqnarray}
where $T_{\gamma}$ is the present photon temperature, $\rho_{crit}$ is the critical density. For solutions satisfying (\ref{om-fidm}) see fig. \ref{Boltzman}, right panel.

\subsection{DM Direct Detection}

\begin{figure}[tp]
  \centering
\includegraphics[width= 8 cm, height = 4 cm]{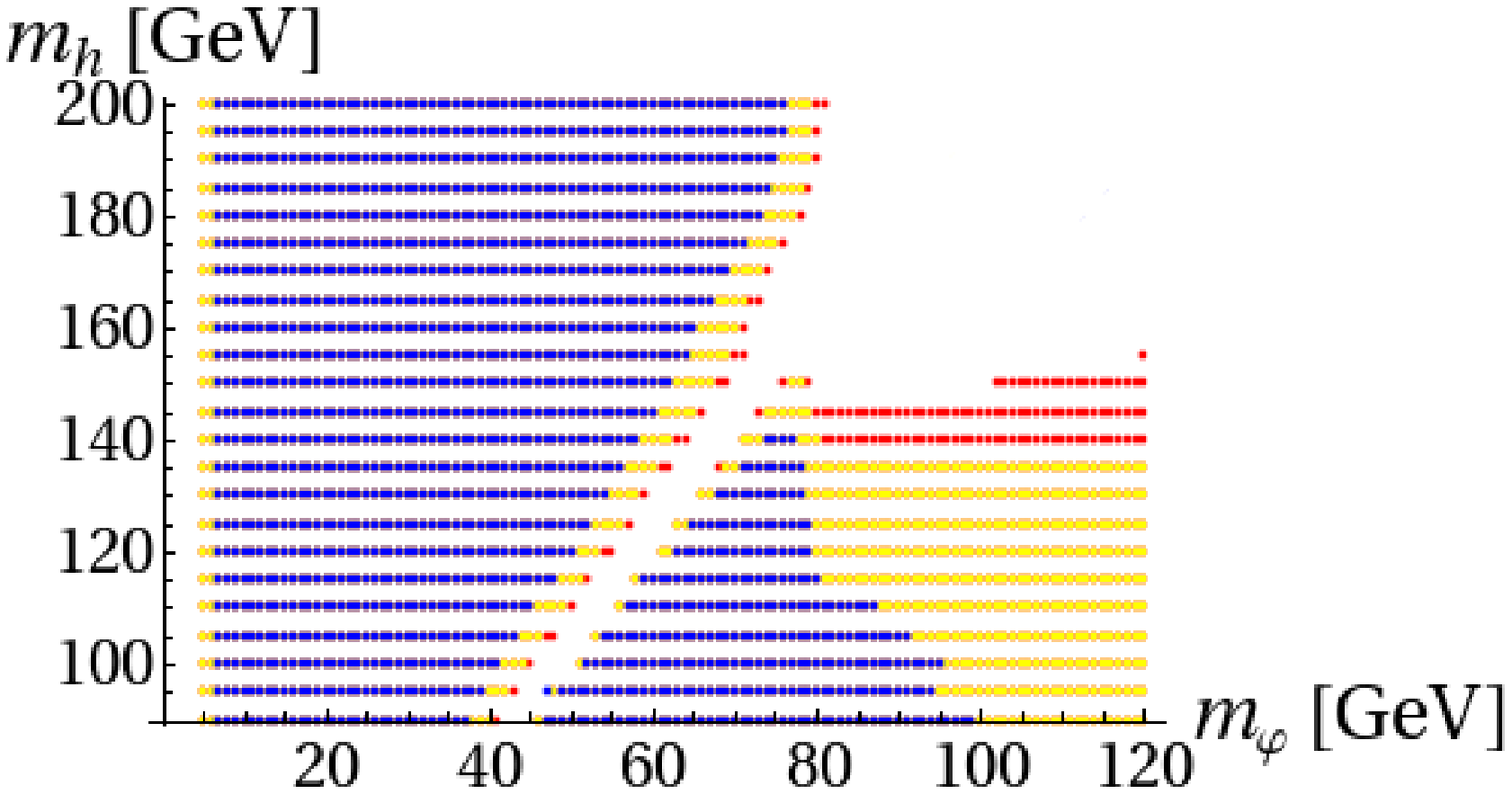}\,\,\,\,
\includegraphics[width=3.7 cm, height = 4 cm]{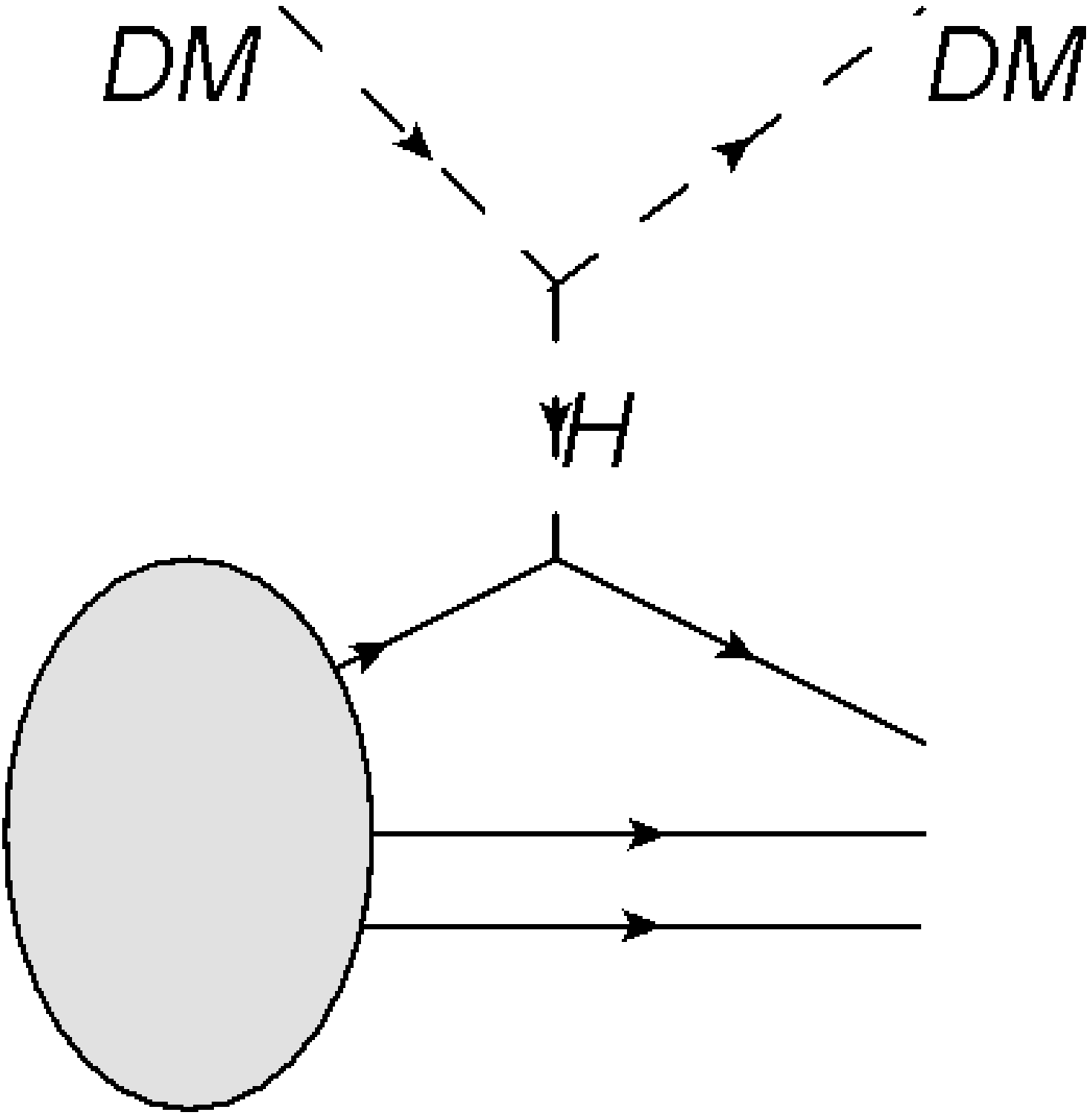}
  \caption{LEFT: XENON100 constraint on $(m_h, m_{\varphi})$ combined with DM abundance in the case of CDM (blue region forbidden for $N=1$, blue and yellow for $N=6$, blue, yellow and red for $N=12$).
RIGHT: Elastic scattering of $\vec{\varphi}$ off a nucleon.}
  \label{directXENON}
\end{figure}

The direct detection rate of $\vec{\varphi}$ is determined by the cross section of $\vec{\varphi}$ scattering off nuclei (see the Feynman diagram in fig.~\ref{directXENON}), which can be found e.g. in \cite{Guo:2010hq} for $N=1$ and in \cite{plans} for compound DM case.
 In fig.~\ref{directXENON} we show allowed regions in the $(m_h, m_{\varphi})$ plane that remain after imposing  limits on the elastic scattering of DM particles off nucleons from XENON100 (the strongest limits on $\sigma_{DM \, N \to DM \, N}$ in the mass range of our interest, \cite{Aprile:2011hi}). The white band corresponds to the resonance region seen in fig.~\ref{phases} for $m_h \sim 2 m_{\varphi}$ in which the annihilation is amplified and the coupling $\lambda_x$ is suppressed. Since the XENON100 data start at $m_{\varphi}=5$ GeV therefore the vertical strip of masses below 5 GeV is also allowed.

\subsection{Self-Interacting DM}\label{sec:SIDM}

\begin{figure}[tp]
  \centering
\includegraphics[height= 4. cm]{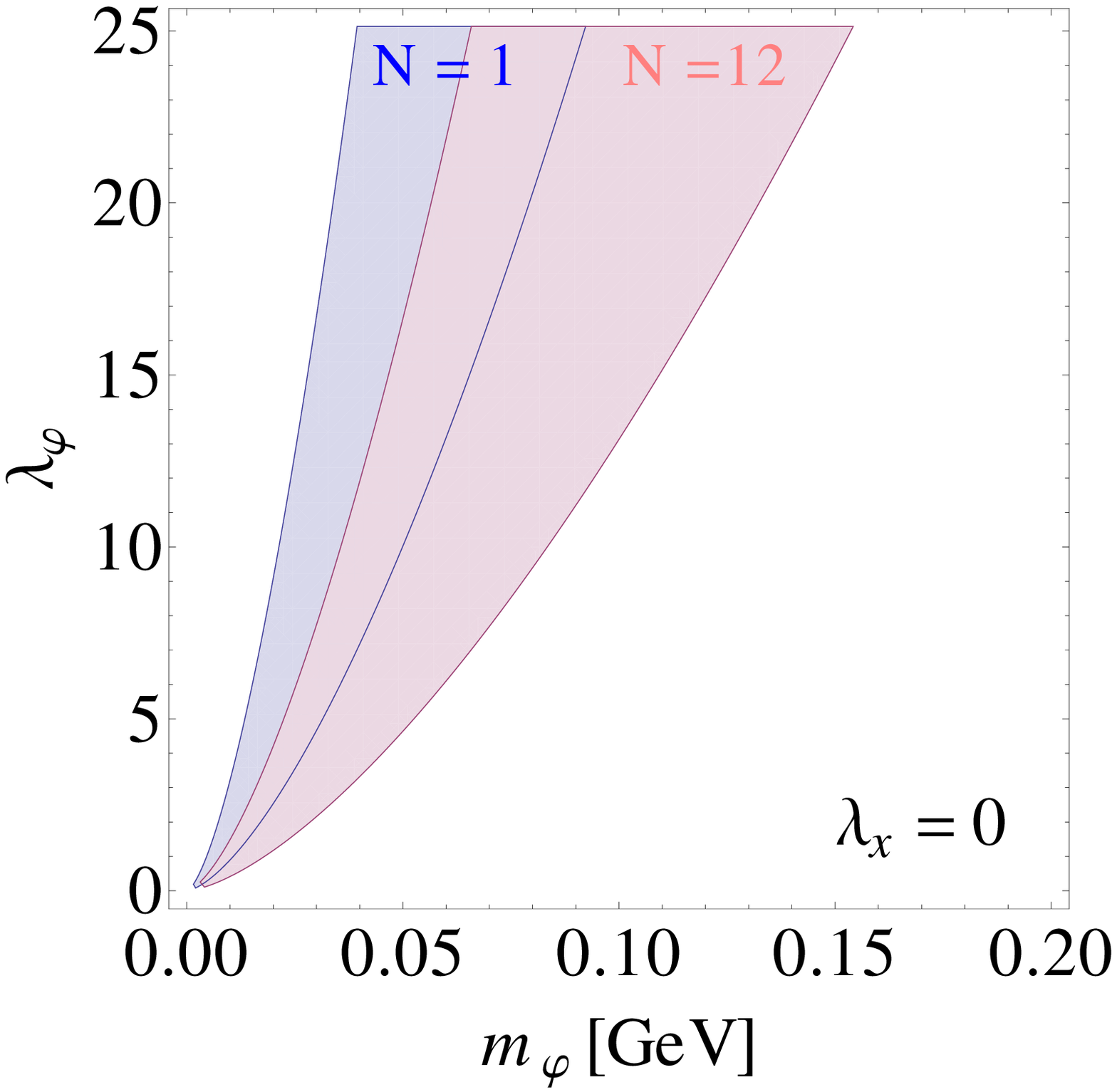}
\includegraphics[height= 4. cm]{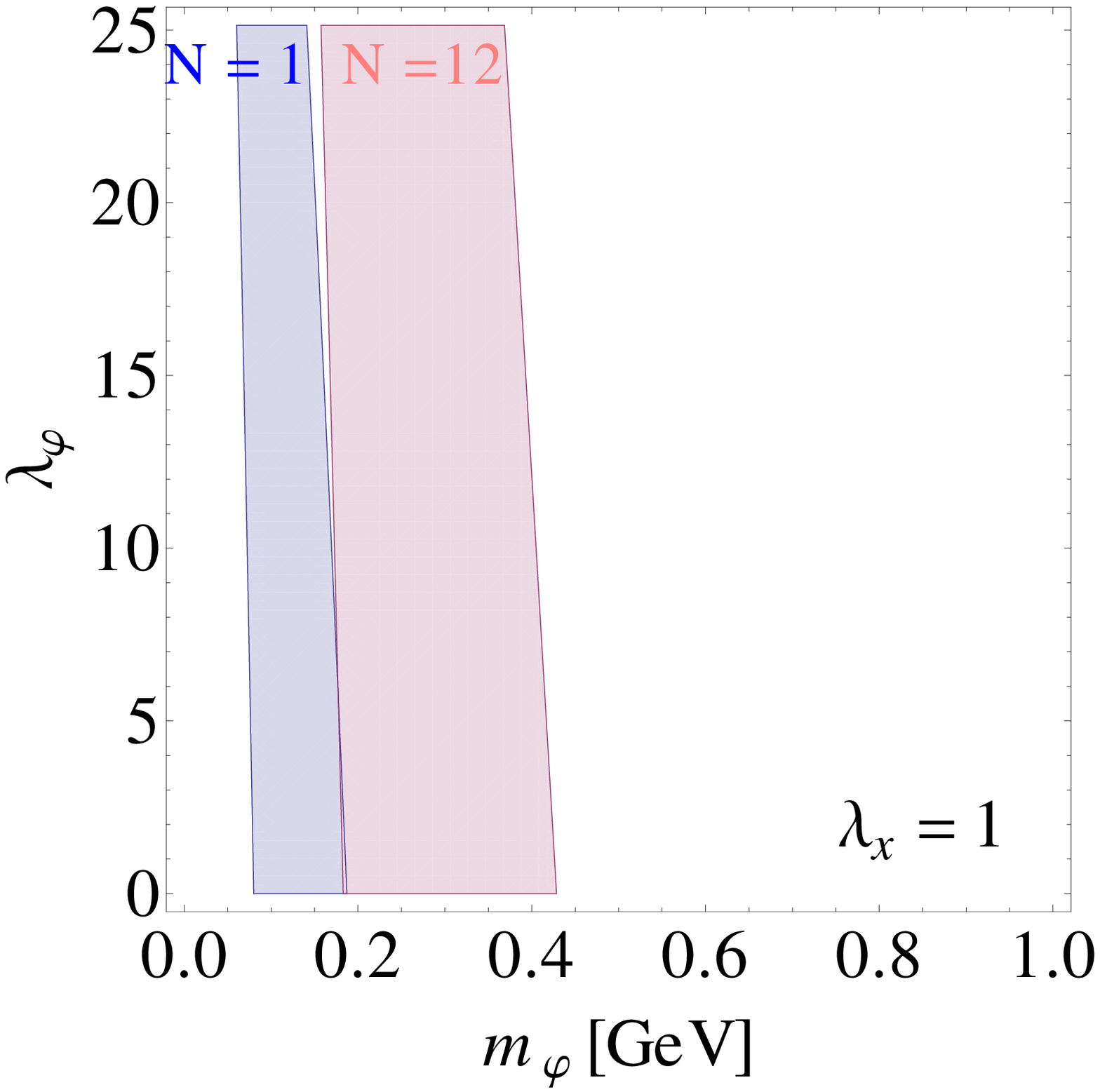}
  \caption{Regions in $(\lambda_{\varphi}, m_{\varphi})$ space allowed by the Steinhardt \& Spergel constraint for $N=1$ (darker region) and $N=12$ (lighter region) for $\lambda_x = 0$ (left), 1 (right).}
\label{SS}
\end{figure} 
The standard  $\Lambda$CDM model is facing some difficulties if compared with observations. High-resolution N-body simulations have shown that the model generates cusps of the DM density distribution in central regions of galaxies~\cite{Navarro:1996gj} and the number of subhalos predicted by the model turns out to be larger than the observed~\cite{Klypin:1999uc} number. Self-interacting DM (SIDM) was proposed by Spergel \& Steinhardt~\cite{Spergel:1999mh} to cure those problems.  

The key feature of SIDM is that the mean free path of DM particles should be between 1 kpc and 1 Mpc in regions where the dark matter density is about $0.4 \mathrm{\,GeV}/{\mathrm cm}^3$. In terms of the unit mass cross section the Steinhardt \& Spergel hypothesis requires that:
\begin{equation}
2.05 \cdot 10^3 \,\mathrm{ GeV}^{-3} \leq \sigma_{DM+DM \to DM+DM} / m_{DM} \leq 2.57 \cdot 10^4 \,\mathrm{ GeV}^{-3}
\label{SS_con}
\end{equation}
For different $N$ this condition implies a relation between $\lambda_x, \lambda_{\varphi}$ and $m_{\varphi}$ illustrated in fig.~\ref{SS} as an allowed region in the ($ \lambda_{\varphi},m_{\varphi}$) space. In the plots $\lambda_{\varphi}$ varies form 0 up to its maximal value allowed by unitarity, i.e. $8\pi$. Similar results were obtained in other versions of scalar DM models \cite{Bento:2001yk}, \cite{McDonald:2001vt}.
As observed from fig.~\ref{phases},  $m_{\varphi}$ consistent with the Spergel and Steinhardt condition is so small that it is not compatible with the CDM case. The only viable option is the FIDM, therefore $\lambda_x \sim 0$ (only the first panel in fig.~\ref{SS} is consistent with the DM abundance). In this case, from fig.~\ref{Boltzman} one can see that $m_{\varphi} \sim 0.01-0.15$ GeV corresponds to $\lambda_x \sim 10^{-10}-10^{-12}$ for $N=1-12$. 

\section{Summary and conclusions}
\label{Sec:sum}

We have considered an extension of the Standard Model by an addition of $N$ real scalar
singlets $\varphi$ with $O(N)$ symmetry that are candidates for Dark Matter. We have discussed theoretical and experimental (cosmological) constraints on the model parameters. The XENON100 direct DM detection experiment gives no constraints on the model in the FIDM case (too small coupling of DM to the SM), but constrain strongly the CDM solution (see fig.~\ref{directXENON}). We have shown that Steinhardt \& Spergel solution of the DM density distribution problem within the singlet scalar SM extension requires feebly interacting DM.

\section{Acknowledgments}
This work is supported in part by the Ministry of Science and Higher Education (Poland) as research project N~N202~006334 (2008-11). 
The authors thank the XENON collaboration for providing relevant data that were adopted for plots shown in this work. AD acknowledges financial support from the project "International PhD Studies in Fundamental Problems of  Quantum Gravity and Quantum Field Theory" of Foundation for Polish Science, cofinanced from the programme IE OP 2007-2013 within European Regional Development Fund.

\end{document}